\documentclass[openacc]{rsproca_new}
\usepackage{bm}

\newcommand{\bmb}{\bm b}
\newcommand{\bmq}{\bm q}
\newcommand{\bmu}{\bm u}

\newcommand\calP{{\cal P}}
\newcommand\calS{{\cal S}}
\newcommand\calT{{\cal T}}
\newcommand\calU{{\cal U}}

\newcommand{\bmzhat}{\hat {\bm z}}

\newcommand{\bmB}{\bm B}

\newcommand{\bmnabla}{\bm \nabla}

\begin{document}

\title{The convective instability of a Maxwell-Cattaneo fluid in the presence of a vertical magnetic field}

\author{
I.A. Eltayeb$^{1}$, D.W. Hughes$^{2}$ and M.R.E.~Proctor$^{3}$}

\address{$^{1}$Department of Mathematical and Physical Sciences, College of Arts and Science, University of Nizwa, Birkat Al Mouz, Nizwa 616, Oman \\
$^{2}$School of Mathematics, University of Leeds, Leeds LS2 9JT, UK \\
$^{3}$DAMTP, Centre for Mathematical Sciences, Wilberforce Road, Cambridge CB3 0WA, UK}

\subject{applied mathematics, fluid mechanics, mathematical modelling}

\keywords{Rayleigh-B\'enard convection, magnetoconvection, hyperbolic heat flow}

\corres{D.W.\ Hughes\\
\email{d.w.hughes@leeds.ac.uk}}

\begin{abstract}

We study the instability of a B\'enard layer subject to a vertical uniform magnetic field, in which the fluid obeys the Maxwell-Cattaneo (MC) heat flux-temperature relation. We extend the work of Bissell (\textit{Proc. R. Soc. A} \textbf{472}: 20160649, 2016) to non-zero values of the magnetic Prandtl number $p_m$. With non-zero $p_m$, the order of the dispersion relation is increased, leading to considerably richer behaviour. An asymptotic analysis at large values of the Chandrasekhar number $Q$ confirms that the MC effect becomes important when $C  Q^{1/2}$ is $O(1)$, where $C$ is the Maxwell-Cattaneo number. In this regime, we derive a scaled system that is independent of $Q$. When $CQ^{1/2}$ is large, the results are consistent with those derived from the governing equations in the limit of Prandtl number $p\to \infty$ with $p_m$ finite; here we identify a new mode of instability, which is due neither to inertial nor induction effects. In the large $p_m$ regime, we show how a transition can occur between oscillatory modes of different horizontal scale. For $Q \gg 1$ and small values of $p$, we show that the critical Rayleigh number is non-monotonic in $p$ provided that $C>1/6$. While the analysis of this paper is performed for stress-free boundaries, it can be shown that other types of mechanical boundary conditions give the same leading order results.

\end{abstract}


\begin{fmtext}
\end{fmtext}
\maketitle

\section{Introduction}
The recent interest in the dynamics of Maxwell-Cattaneo (or non-Fourier) fluids is motivated by a variety of  practical applications for which the Fourier law of heat flux is inadequate. Maxwell \cite{Maxwell_1867}, in his study of the theory of gases, proposed that the relation between the heat flux and the temperature gradient should not be instantaneous (as indeed is disallowed by relativity theory), but instead must involve a finite relaxation time. Cattaneo \cite{Cattaneo_1948} proposed a similar relation for solids, which was developed further by Oldroyd \cite{Oldroyd_1950}. Other important contributions were made later, by, for example, Fox \cite{Fox_1969} and Carrassi \& Morro \cite{CM_1972}. 

The idea of a finite relaxation time is incorporated into the Maxwell-Cattaneo (MC) relation between the heat flux $\bmq$ and the temperature $T$, which takes the form
\begin{equation} 
\tau_r \frac{\mathrm{D} \bmq}{\mathrm{D} t} = - \bmq - K \bmnabla T, 
\label{eq:MC_flux}
\end{equation}
in which $\tau _r$ is the relaxation time and $K$ is the thermal conductivity. The introduction of a finite relaxation time changes the fundamental nature of the parabolic heat equation of Fourier fluids, in which heat diffuses with infinite speed, to a hyperbolic heat equation with the solution of a heat wave that propagates with finite speed \cite{JP_1989, Straughan_book_2011}. When $\tau_r=0$, we recover the Fourier law, with an instantaneous flux-gradient relation. The importance of the thermal relaxation term is typically expressed via the Maxwell-Cattaneo coefficient $C$, which is defined as the ratio of the thermal relaxation time to twice the thermal diffusion time; i.e.\ $C = \tau_r K / (2 \rho c_p d^2)$, where $\rho$ is the density, $c_p$ is the specific heat at constant pressure and $d$ is a representative length scale. For later use, the thermal diffusivity $\kappa$ is defined as $\kappa = K/(\rho c_p)$. Thus the classical Fourier law has $C=0$.

The Maxwell-Cattaneo heat transport effect has been studied in a wide variety of different physical contexts: for example, in solids \cite{BZ_1997}, in fluids \cite{SF_1984, LC_1984, Straughan_2009, Straughan_2010, Stranges_etal_2013, Bissell_2015}, in porous media \cite{Straughan_2013}, in nanofluids and nanomaterials \cite{Jou_etal_2011, Lebon_etal_2011}, in liquid helium \cite{LL_1984, Donnelly_2009}, in biological tissues \cite{Dai_etal_2008, Tung_etal_2009}, and, in the context of double diffusive convection, in stellar interiors \cite{HF_1995, Gnedin_etal_2001, Straughan_2011}. 

The problem of (non-magnetic) convection incorporating MC heat transport has received considerable attention  \cite{SF_1984, Straughan_2009, Straughan_2010, Stranges_etal_2013, Bissell_2015}. Whereas standard Rayleigh-B\'enard convection is susceptible only to steady (direct) instability (so-called `exchange of stabilities') \cite{Chandra_1961}, the MC effect can also lead to instability occurring in an oscillatory manner; indeed, this is the preferred mode if $C$ is sufficiently large. The accuracy of the Fourier law for experiments involving simple classical fluids suggests that $\tau_r$ is usually very small, as shown in \cite{CM_1972}, where it is suggested that $\tau_r$ can be as small as $10^{-9} \mathrm{s}$ for gases; this leads to very small values of $C$. However, for more complex fluids, $\tau_r$, and hence $C$, can be much larger.  For example, Neuhauser \cite{Neuhauser_1987}, in her investigation of thermal relaxation in superfluid helium-3, found that $\tau_r$ has values in the range 30s -- 400s; Mohammadein \cite{Mohammadein_2006}, in his study of the thermal relaxation time in two-phase bubbly flow, found that the relaxation time varied between $10^{-3}$s and 3s. It is though important to bear in mind that even though $C$ may be extremely small, passing from $C=0$ to $C\ll 1$ represents a singular perturbation, which permits new types of solution that involve $C$ in an essential way. 

In this paper, motivated by astrophysical applications, we consider in detail the role of the MC effect in the onset of convection in an imposed vertical magnetic field (magnetoconvection). In the classical ($C=0$) problem, instability can occur either as a steady or oscillatory mode, depending on the values of the various governing parameters \cite{WP_2014}. It is therefore of interest to understand how the stability properties of oscillatory modes are influenced by the MC effect, which itself leads to oscillations being preferred. Also, as we shall explain in detail below, very high magnetic field strengths, as can occur astrophysically, can lead to the MC term having a significant effect on the dynamics even when $C$ is tiny.

The linear stability problem for magnetoconvection with the MC effect has been analysed by Bissell \cite{Bissell_2016}, who, however, considered only the case of zero magnetic Prandtl number ($p_m = \nu/\eta$, where $\nu$ is the kinematic viscosity and $\eta$ the magnetic diffusivity; compare the Prandtl number $p=\nu/\kappa$, where $\kappa$ is the thermal conductivity). While this is appropriate for terrestrial situations, in which $p_m$ is very small, in astrophysical situations, $p_m$ can be $O(1)$ (stellar interiors) or larger (interstellar medium). Indeed, even for the classical magnetoconvection problem ($C=0$), much of the interesting dynamics arises only for $p_m>p$. The extension to non-zero $p_m$ is non-trivial, since, whereas for the $p_m=0$ case the linear system is governed by a third order dispersion relation, as is the classical problem of magnetoconvection, finite $p_m$ introduces a new mode, with a fourth order dispersion relation. In general, the stability problem must be solved numerically. However, it is possible to make analytical progress in a number of limiting cases; in particular, in the case where the Chandrasekhar number $Q$, a measure of the strength of the imposed magnetic field, is large; in an astrophysical context, this is an important and well-studied regime for the classical problem. Here we are able to show, via a precise asymptotic ordering, that the influence of the MC effect is felt strongly (i.e.\ at leading order) for very small values of $C$, with $C=O(Q^{-1/2})$.

The layout of the paper is as follows. Section~2 contains the mathematical formulation of the problem. The main stability results are contained in \S\,3; in \S\,3(a) we derive the governing dispersion relation; the instability criteria are derived in \S\,3(b); numerical results for various limiting cases are contained in \S\,3(c). Section~4 considers the role of $p_m$ in determining the preferred behaviour. Section~5 considers the large $Q$ regime, and the paper concludes with a discussion in \S\,6 of the astrophysical implications.

\section{Formulation of the problem}
We consider a horizontal layer of an incompressible (Boussinesq) viscous Maxwell-Cattaneo fluid, initially at rest, contained between two planes a distance $d$ apart. A uniform vertical magnetic field $\bmB = B_0 \bmzhat$ is imposed. The lower plane ($z=0$) and the upper plane ($z=d$) are maintained at uniform temperatures $T_B$, $T_T$, respectively, with $T_B > T_T$, so that the layer is heated from below, giving rise to a uniform adverse temperature gradient $\hat \beta$. This basic state is given a small disturbance, giving rise to perturbations in velocity $\bmu$, pressure $p$, temperature $\theta$, density $\rho$, magnetic field $\bmb$ and heat flux $\bmq$. On introducing characteristic units of distance, time, velocity, heat flux, temperature, magnetic field and pressure by $d$, $d^2/\kappa$, $\kappa /d$, $\hat \beta K$, $\hat \beta d$, $B_0 \eta/\nu$ and $\rho _0 \nu \kappa /d^2$, the dimensionless linearised perturbation equations can be written as 
\begin{equation} 
\frac{1}{p} \frac{\partial \bmu}{\partial t} = -\bmnabla \Pi + R \theta \bmzhat + p Q \frac{\partial \bmb}{\partial z} + \nabla ^2 \bmu ,
\label{eq:mom}
\end{equation}
\begin{equation} 
\frac{\partial \theta }{\partial t} = \bmu \cdot \bmzhat-F,
\label{eq:temp}
\end{equation}
\begin{equation}
2C \frac{\partial F}{\partial t}+ F+ \nabla ^2 \theta =0,
\label{eq:flux}
\end{equation}
\begin{equation} 
p_m \frac{\partial \bmb}{\partial t} = \frac{\partial \bmu}{\partial z} + p \nabla ^2 \bmb,
\label{eq:ind}
\end{equation}
\begin{equation} 
\nabla \cdot \bmu=\nabla \cdot \bmb=0,
\label{eq:sol}
\end{equation}
where $\Pi $ is the total pressure (gas $+$ magnetic) and $F=\bmnabla \cdot \bmq$. The Rayleigh number $R$, Chandrasekhar number $Q$, Maxwell-Cattaneo coefficient $C$, Prandtl number $p$ and magnetic Prandtl number $p_m$ are defined as
\begin{equation} 
R= \frac{g \hat \alpha \hat \beta d^4}{\kappa \nu}, \quad
Q=\frac{B_0^2 d^2}{\mu_0 \rho_0 \eta \nu}, \quad
C= \frac{\tau_r \kappa}{2 d^2}, \quad
p=\frac{\nu}{\kappa}, \quad
p_m=\frac{\nu}{\eta}, \quad
\label{eq:pars}
\end{equation}
where $\tau_r$ is the thermal relaxation time, $\eta$ is the magnetic diffusivity, $\rho_0$ is the reference density, $\mu_0$ is the permeability and $\hat \alpha$ is the coefficient of thermal expansion. Note that the scaling of the magnetic field has been chosen in order that the limit of $p_m \to 0$ is straightforward. The equations can also be formulated using $\zeta = \eta/\kappa$ rather than $p_m$ (see, for example, \cite{WP_2014}).

Equations \eqref{eq:mom} -- \eqref{eq:sol} are solved subject to boundary conditions depending on the thermal, dynamical and electrical properties of the boundaries. We define a Cartesian coordinate system $(x,y,z)$ with the fluid confined to the (dimensionless) region $0<z<1$. All variables are periodic in the horizontal directions. The horizontal boundaries are perfectly thermally conducting, impermeable and stress-free, leading to the conditions
\begin{equation} 
\theta = u_z = \frac{\partial u_x}{\partial z} = \frac{\partial u_y}{\partial z} = 0 \quad \textrm{at} \quad z=0, 1,
\label{eq:vel_sf}
\end{equation}
where $\bmu = (u_x, u_y, u_z)$. We discuss other mechanical boundary conditions in \S\,6.

The magnetic boundary conditions are more complicated, since they depend on the electrical conductivity and magnetic permeability of the boundaries (see, e.g., \cite{RZ_2000, ER_2013}). However, studies of similar problems have shown that the general stability properties are qualitatively similar for different conditions \cite{Eltayeb_1975}. For simplicity, here we follow the counsel of Weiss and Proctor \cite{WP_2014} and adopt the most mathematically tractable conditions, namely  
\begin{equation}
b_x = b_y = \frac{\partial b_z}{\partial z} = 0 \quad \textrm{at} \quad z=0, 1,
\label{eq:mag_bc}
\end{equation}
where $\bmb = (b_x, b_y, b_z)$.

\section{Stability analysis}

\subsection{The dispersion relation}

It is convenient to decompose the solenoidal fields $\bmu$ and $\bmb$ into their poloidal and toroidal parts by writing
\begin{equation}
\bmu = \bmu_P + \bmu_T, \quad
\bmu_P= \bmnabla \times \bmnabla \times \calP \bmzhat, \quad
\bmu_T= \bmnabla \times \calT \bmzhat,
\end{equation}
\begin{equation}
\bmb = \bmb_P + \bmb_T, \quad
\bmb_P= \bmnabla \times \bmnabla \times \calS \bmzhat, \quad
\bmb_T= \bmnabla \times \calU \bmzhat.
\end{equation}
As in the classical magnetoconvection problem (see, for example, \cite{WP_2014}), the equations for $\calT$ and $\calU$ describe decaying Alfv\'en waves, which are not coupled to the convection. Thus we need concentrate only on the equations for $\calP$ and $\calS$. The governing equations are therefore
\begin{align}
\label{eq:goveq1}
\frac{1}{p} \frac{\partial}{\partial t} \left( \nabla^2 \calP \right) &= - R \theta + p Q \frac{\partial}{\partial z} \left( \nabla^2 \calS \right) + \nabla^4 \calP, \\
\label{eq:goveq2}
\left( 2 C \frac{\partial}{\partial t} + 1 \right) \left(\frac{\partial \theta}{\partial t} + \nabla_H^2 \calP \right) &= \nabla^2 \theta,\\
\label{eq:goveq3}
p_m \frac{\partial \calS}{\partial t} &= \frac{\partial \calP}{\partial z} + p \nabla^2 \calS,
\end{align}
where $\nabla_H^2$ is the horizontal Laplacian. Equation~\eqref{eq:goveq1} is derived from the $z$-component of the curl of the curl of the momentum equation~\eqref{eq:mom}; equation~\eqref{eq:goveq2} comes from combining equations~\eqref{eq:temp} and \eqref{eq:flux}; equation~\eqref{eq:goveq3} is derived from the $z$-component of the induction equation~\eqref{eq:ind}.

As in the classical problem, for the boundary conditions given by \eqref{eq:temp_bc} -- \eqref{eq:mag_bc}, we seek normal modes with
\begin{equation}
\theta \propto \calP \propto f(x,y) \sin \pi z \, e^{st}, \quad
\calS \propto f(x,y) \cos \pi z \, e^{st},
\label{eq:normal_modes}
\end{equation}
where the planform function $f(x,y)$ satisfies
\begin{equation}
\nabla_H^2 f = -k^2 f.
\end{equation}
Substituting from \eqref{eq:normal_modes} into equations~\eqref{eq:goveq1} -- \eqref{eq:goveq3} leads to the dispersion relation
\begin{align} \nonumber
\beta^2 \left(s(2Cs+1)+\beta^2\right) (s+ p \beta^2)(p_m s+p \beta^2) - p R k^2 &(2Cs+1)(p_m s+ p\beta^2) \\
 &+ p^2 Q \pi^2 \beta^2 \left(s(2Cs+1)+\beta^2\right)=0,
\label{eq:dr1}
\end{align}
where $\beta^2=k^2+\pi^2$. This expression reduces, as it must, to expression~(3.26) in \cite{WP_2014} when $C=0$, and to expression~(4.3) in \cite{Bissell_2015} when $Q=0$. We remove all factors of $\pi^2$ by the substitution
\begin{equation}
k = \pi {\tilde k}, \quad \beta= \pi {\tilde \beta}, \quad s=\pi^2 {\tilde s}, \quad C = {\tilde C}/\pi^2, \quad R= \pi^4 {\tilde R}, \quad Q= \pi^2 {\tilde Q}.
\end{equation}
On dropping the tildes, \eqref{eq:dr1} then becomes
\begin{align} \nonumber
\beta^2 \left(s(2Cs+1)+\beta^2\right) (s+ p \beta^2)(p_m s+p \beta^2) - p R k^2 &(2Cs+1)(p_m s+ p\beta^2) \\
 &+ p^2 Q \beta^2 \left(s(2Cs+1)+\beta^2\right)=0,
\label{eq:dr2}
\end{align}
where now $\beta^2= k^2+1$. This is a quartic polynomial in $s$:
\begin{equation}
a_4s^4 + a_3s^3 + a_2s^2 +a_1s + a_0=0,
\label{eq:dr3}
\end{equation}
with
\begin{equation}
a_4 = 2Cp_m,
\end{equation}
\begin{equation}
a_3 = p_m + 2C p (1+p_m) \beta^2 ,
\end{equation}
\begin{equation}
a_2 = (p+p_m+p p_m) \beta^2 + 2C p^2 (\beta^4 + Q) - 2C p p_m R k^2 / \beta^2,
\label{eq:a2}
\end{equation}
\begin{equation}
a_1 = (p + p^2 + p p_m) \beta^4 + p^2 Q - p p_m R k^2 / \beta^2 - 2C p^2 R k^2 ,
\end{equation}
\begin{equation}
a_0 = p^2 (\beta^6 + Q \beta^2 - R k^2).
\label{eq:a0}
\end{equation}

\subsection{Bifurcations to instability}

Bifurcations to instability can occur in either a steady or oscillatory fashion. Steady bifurcations occur when $s=0$, with the value of $R$ given by
\begin{equation}
R = R^{(s)} = \frac{\beta^6 + Q\beta^2}{k^2}.
\label{eq:steady}
\end{equation}
It is important to note that there is no dependence of $R^{(s)}$ on $p$ (provided that it is non-zero), $p_m$ or, particularly, $C$. Thus the MC effect has no influence on the onset of steady convection; criterion~\eqref{eq:steady} is simply that of the classical problem \cite{Chandra_1961, WP_2014}. It can be seen from \eqref{eq:steady} that $R^{(s)}$ is minimised at a finite value of $k^2$, $k^2_{sc}$, say; we denote this minimum value of $R^{(s)}$ by $R_c^{(s)}$.

To determine the occurrence of oscillatory bifurcations, we set $s=i \omega$, with $\omega$ real. Taking the real and imaginary parts of the dispersion relation~\eqref{eq:dr3} leads to the equations:
\begin{equation}
\label{eq:omega_two_exp}
a_4 \omega^4 - a_2 \omega^2 + a_0 = 0, \qquad \omega^2 = a_1/a_3,
\end{equation}
which yield two expressions for $\omega^2$ and $R$, where $\omega^2$ must be positive at a point of bifurcation. Eliminating $R$ leads to a quadratic expression for $\omega^2$, whereas eliminating $\omega^2$ leads to a quadratic expression for $R$; both turn out to be useful. The equation for $\omega^2$ is
\begin{equation}
b_4 \omega^4 + b_2 \omega^2 + b_0 =0,
\label{eq:omega}
\end{equation}
where
\begin{equation}
 b_4= 4 C^2 p p_m^2,
\end{equation}
\begin{equation}
 b_2 = p_m^2 (1 + p - 2 C p \beta^2) + 4 C^2 p^3 ( \beta^4 + Q),
\end{equation}
\begin{equation}
 b_0 = p^2 (1 + p) \beta^4 + p^2 (p-p_m) Q - 2 C p^3 (\beta^4+Q) \beta^2.
\end{equation}
The equation for $R$ is
\begin{equation}
c_2 R^2 + c_1 R + c_0 =0,
\label{eq:sb_osc}
\end{equation}
where
\begin{equation}
c_2 = 4 C^2 p^2 p_m^2 (p_m+2Cp \beta^2) ,
\end{equation}
\begin{align}
\nonumber
c_1 &= - \left(8 C^3 p^4 (1+p_m) \left( \frac{\beta^8}{k^2} + Q \frac{\beta^4}{k^2} \right) + 4 C^2 p^2 p_m \left((2 p p_m + p + p_m^2 + p_m) \frac{\beta^6}{k^2} + Q p (1+2p_m)\frac{\beta^2}{k^2} \right) \right. \\
&\qquad \qquad \qquad \qquad \qquad \qquad 
\left. + 2 C  p p_m^2 (1+ p + p p_m)\frac{\beta^4}{k^2} + p_m^3 (1+p)\frac{\beta^2}{k^2} \right),
\end{align}
\begin{align}
\nonumber 
c_0 &= (1+p_m) \left(4 C^2 p^4 (Q+\beta^4)^2 \frac{\beta^4}{k^4} + 
2 C Q p^2 (p p_m + p - 2 p_m) \frac{\beta^6}{k^4} + \right. \\
& \qquad \qquad
 \left. 2 C (p^3 (1+p_m) + p^2 (1+ p_m^2))\frac{\beta^{10}}{k^4}+ 
p^2 p_m Q \frac{\beta^4}{k^4} + p_m(1+p)(p_m+p) \frac{\beta^8}{k^4} \right).
\label{eq_c0}
\end{align}
We denote the smallest value of $R$ given by \eqref{eq:sb_osc}, with $\omega^2>0$, by $R^{(o)}$ (equivalently this can be calculated from either of the expressions in \eqref{eq:omega_two_exp} once $\omega^2>0$ has been determined). There are two possibilities: either $R^{(o)}(k^2)$ has a true minimum $R_c^{(o)}$, with the corresponding value of $k^2$ denoted by $k^2_{oc}$ and the corresponding value of $\omega^2$ by $\omega_c^2> 0$; or the smallest value of $R^{(o)}$ occurs when $\omega=0$ and the oscillatory branch joins the steady branch. The overall critical Rayleigh number at the onset of instability, which we denote by $R_c$, is then given by the minimum value of $R_c^{(s)}$ and $R_c^{(o)}$; we denote the value of $k^2$ at $R=R_c$ by $k_c^2$.

\subsection{Limiting cases}

It is instructive first to recover the special simpler cases of $C=0$, $p_m=0$, $Q=0$, all of which have been studied previously \cite{WP_2014, Bissell_2016, Bissell_2015}.

\subsubsection{$C=0$}

If $C=0$, the coefficient $b_4$ becomes zero; the system then reduces to classical magnetoconvection; there is only one possible mode of oscillation, with
\begin{equation}
\omega^2 = \frac{p^2}{p_m^2}\left( \frac{Q(p_m-p)}{1+p}-\beta^4 \right).
\end{equation}
Noting that $\beta^2 > 1$, we see that for $\omega^2 >0$ we must necessarily have
\begin{equation}
p_m>p \quad \textrm{and} \quad Q> \frac{1+p}{p_m-p}.
\label{eq:magosc}
\end{equation}
In this case, the Rayleigh number for the onset of oscillatory instability is given by
\begin{equation}
R = R^{(o)} = \frac{(p+p_m)(1+p_m)}{p_m^2} \frac{\beta^6}{k^2} + \frac{p^2(1+p_m)}{p_m^2(1+p)} Q \frac{\beta^2}{k^2}.
\label{eq:R_osc}
\end{equation}
If inequalities~\eqref{eq:magosc} are satisfied, then oscillatory instability will be preferred if $R_c^{(o)}$, the minimum of $R^{(o)}$ over all wavenumbers, is less than $R_c^{(s)}$; this is discussed in more detail in \cite{WP_2014}.

\subsubsection{$p_m=0$}

Similarly, if $p_m=0$, then $b_4$ again vanishes, and we recover the system studied in detail in \cite{Bissell_2016}, with
\begin{equation}
\omega^2 = \frac{\beta^2}{2C}-
\frac{\left(pQ + (1 + p)\beta^4\right)}{4 C^2 p ( Q+ \beta^4)}.
\label{eq:omega_pmeq0}
\end{equation}
It is easy to see from \eqref{eq:omega_pmeq0} that $\omega^2$ is positive for sufficiently large $\beta^2$, so oscillations are always possible (for any finite $C$) --- though may not be preferred. Depending on the values of $p$, $Q$ and $C$, $\omega^2$ may be positive for all values of $\beta^2$, or there may be a range of excluded wavenumbers, which does not necessarily include $\beta^2=1$. We shall explore further the case of $p_m=0$ in \S\,\ref{sec:pm}.

\subsubsection{$Q=0$}

For $Q=0$, there is a stable root with $s=-p \beta^2 /p_m$ (most readily seen from expression~\eqref{eq:dr2}). The relation~\eqref{eq:dr3} then becomes a cubic in $s$ and we recover the hydrodynamic Maxwell-Cattaneo problem (see, for example, \cite{Straughan_2010, Bissell_2015}). In this case,
\begin{equation}
\label{eq:omsq_hydro}
\omega^2 = \frac{2Cp\beta^2-(1+p)}{4C^2p},
\end{equation}
with $\omega^2>0$ if
\begin{equation}
\label{eq:beta_bound}
\beta^2 > \frac{1+p}{2Cp}.
\end{equation}
The Rayleigh number for the onset of oscillatory instability is then given by
\begin{equation}
\label{eq:Ro}
R^{(o)}=\frac{1}{2C} \frac{\beta^4}{k^2} + \frac{(1+p)}{4C^2p^2} \frac{\beta^2}{k^2},
\end{equation}
which is minimised when
\begin{equation}
k^2 = k_{oc}^2 = \left( 1 + \frac{(1+p)}{2Cp^2} \right)^{1/2} .
\label{eq:k4}
\end{equation}
As noted in \cite{Bissell_2015}, there is considerable simplification when this expression for $k^2$ is substituted into the expression for $R^{(o)}$, giving
\begin{equation}
R^{(o)}_c = \frac{1}{2C} \left(1+ \left[ 1 + \frac{(1+p)}{2Cp^2}\right]^{1/2} \right)^2.
\label{eq:Ro2}
\end{equation}
Various limiting values of $C$ and $p$ can be explored analytically. As $C \to \infty$,
\begin{equation}
R^{(o)}_c \to \frac{2}{C},
\end{equation}
with $k^2_{oc} = 1 + O(C^{-1})$. Thus, for large $C$, oscillatory convection is preferred since, in this hydrodynamic case, $R^{(s)}_c =27/4$, independent of the value of $C$. \\
As $C \to 0$, there is no minimum on the oscillatory branch; the smallest value of $R^{(o)}$ comes from setting $\omega=0$. Rather than \eqref{eq:Ro2}, this gives, in the limit $C\to 0$,
\begin{equation}
R^{(o)}_c \to \frac{(1+p)^2}{4C^2p^2},
\end{equation}
which is the corrected version of expression~(4.12a) in \cite{Bissell_2015}. Clearly, steady convection is preferred in the limit of small $C$.

As $p \to \infty$, $k_{oc}^2 = 1 + O(p^{-1})$ and hence, from~\eqref{eq:omsq_hydro},
\begin{equation}
\omega^2 \to \frac{4C-1}{4C^2}.
\end{equation}
Thus the $R^{(o)}$ curve has a minimum only for $C>1/4$. In this case,
\begin{equation}
R^{(o)}_c \to \frac{2}{C}.
\end{equation}
Oscillations are therefore preferred only if 
\begin{equation}
\label{eq:C827}
R^{(o)}_c < R^{(s)}_c = 27/4, \quad \textrm{i.e.},\ \ C>8/27.
\end{equation}
For $1/4<C<8/27$, the oscillation curve has a minimum but $R^{(o)}_c > R^{(s)}_c$; for $C<1/4$ the oscillation curve has no minimum.\\
As $p \to 0$, the optimal value of $k^2 \to 1/(\sqrt{2C} p)$, which leads to $\omega^2>0$ provided that $C>1/2$, and gives
\begin{equation}
\label{eq:Romin_small_p}
R^{(o)}_c \to \frac{1}{4C^2p^2}.
\end{equation}
For $C<1/2$, $R^{(o)}$ is minimised for the smallest $k^2$ that allows a positive $\omega^2$, i.e.\ $k^2 \approx 1/(2Cp)$. At leading order this again yields the estimate~\eqref{eq:Romin_small_p}. Clearly, regardless of the value of $C$, steady convection is preferred as $p \to 0$.

\medskip

In each of these three limiting cases, the dispersion relation~\eqref{eq:dr3} reduces to a cubic, so there can be at most one mode of oscillation for any values of the parameters. In the general case, however, for any given set of parameters, it is not immediately obvious whether there can be more than one mode of oscillation --- in other words, whether equation~\eqref{eq:omega} can have two roots both with $\omega^2 > 0$. This would require satisfaction of the two conditions
\begin{equation}
b_0>0, \qquad
b_2< - \sqrt{4 b_0 b_4}.
\label{eq:2roots}
\end{equation}
Given the complicated form of the coefficients $b_j$, it is not possible to make enormous analytical progress in determining whether inequalities~\eqref{eq:2roots} are satisfied. However, we have conducted an extensive search over the full $(C, Q, p, p_m, \beta^2)$ parameter space, and have not found any values satisfying the criterion~\eqref{eq:2roots}. Thus there can be no more than one mode of oscillation for any combination of values of the parameters. As long as the parameters remain finite, the oscillatory branch can end only on the stability branch, where the frequency tends to zero.

\section{The role of $p_m$}\label{sec:pm}

We now proceed to identify the dependence of the critical (preferred) mode of convection on the governing parameters, paying particular attention to the role of non-zero $p_m$, thereby extending the results of \cite{Bissell_2016}. For any given values of the parameters, the Rayleigh number for the onset of steady convection, $R^{(s)}$, given by \eqref{eq:steady}, exists for all values of $\beta^2$, with a well-defined minimum. The Rayleigh number for the onset of oscillatory convection, $R^{(o)}$, exists for at least some range of $\beta^2$; interestingly, for the magnetoconvection problem ($C=0$), this is guaranteed for sufficiently small $\beta$, provided that conditions~\eqref{eq:magosc} are satisfied, whereas for any $C>0$, oscillations are guaranteed for sufficiently large $\beta^2$. Extensive computations have been carried out to identify the critical mode in the four-dimensional parameter space ($p, p_m, C, Q$).

\begin{figure}[!h]
\centering\includegraphics[trim=0.5cm 0 0 0cm, width=5.3in]{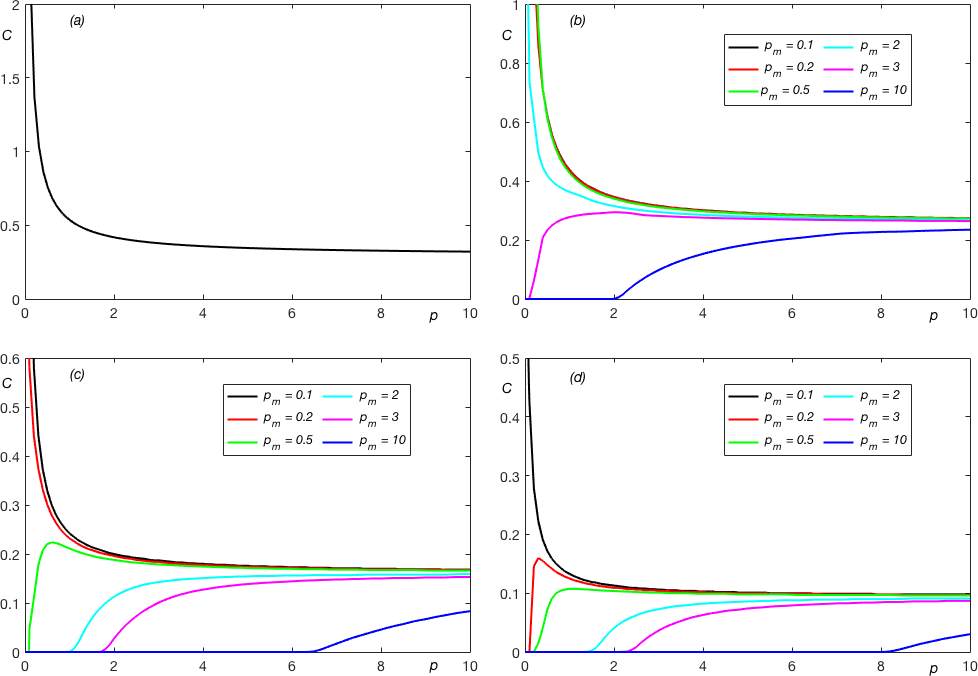}
\caption{Regime diagrams showing the regions of preference of the steady mode (below the curves) and the oscillatory mode (above the curves) in the $\left(p, C\right)$ plane for (a)~$Q=0$, (b)~$Q=1$, (c)~$Q=10$, (d)~$Q=50$ at various marked values of $p_m$. In~(a), the value of $p_m$ is irrelevant; in~(b), the curves for $p_m=0.1$, $0.2$ and $0.5$ essentially coincide.}
\label{fig:fig1}
\end{figure}

In general, the stationary mode is preferred when the parameters $p_m$, $C$, $Q$ are small. However, when any of these parameters is sufficiently large, the preferred mode at onset becomes oscillatory. The salient features of the critical mode are summarised in the regime diagram in figure~\ref{fig:fig1}, which exhibits the regions of preference for stationary and oscillatory modes in the $\left( p, C \right)$ plane for different values of $p_m$ and $Q$ (for $Q=0$, the value of $p_m$ is of course irrelevant). For $Q=0$, the stationary mode is preferred in regions with small values of $C$, while the oscillatory mode is preferred for larger values of $C$. The value of $C$ that marks the transition between steady and oscillatory modes has $C \to 1/(3 \sqrt{3}p)$ for $p \to 0$ and $C \to 8/27$ for $p \to \infty$; both of these regimes are captured in figure~\ref{fig:fig1}$a$. For small values of $Q$, this picture is essentially unchanged, for any $O(1)$ values of $p_m$. As can be seen in in figures~\ref{fig:fig1}$b$--$d$, for any fixed $Q > 0$, the region of preference for the stationary mode shrinks as $p_m$ increases; i.e., increasing the magnetic Prandtl number promotes oscillatory instability. When $p_m$ is kept fixed and $Q$ is increased, the region of preference of the stationary mode is also diminished; indeed, for sufficiently small $p$, oscillatory modes can be preferred for all values of $C$. The change in the preferred mode from stationary to oscillatory is typically accompanied by sudden changes in the wavenumber and frequency; by definition, the critical Rayleigh number curve remains continuous, although its slope may change. 

\begin{figure}[!h]
\centering\includegraphics[trim=1cm 0 0 0cm, width=5in]{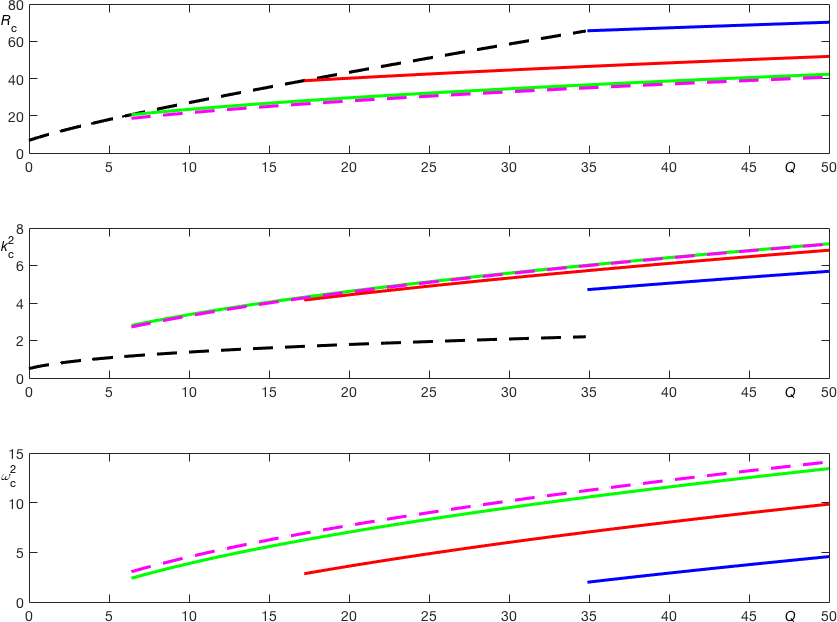}
\caption{$R_c$, $k_c^2$ and $\omega_c^2$ as functions of $Q$ for $p_m=0$ and $C=0.2$, with $p=0.5$ (blue), $p=1$ (red), $p=5$ (green). Bifurcation to oscillatory convection is shown as solid lines; the onset of steady convection, which is independent of $p$, is depicted by a dashed black line. The magenta dashed lines show the asymptotic results~\eqref{eq:large_p_asymp} for large $p$.}
\label{fig:fig2}
\end{figure}

Before exploring the effects of non-zero $p_m$ in detail, it is helpful to revisit the case of $p_m=0$ \cite{Bissell_2016}. Figure~\ref{fig:fig2} presents examples of the change of critical mode from stationary to oscillatory as $Q$ is increased, for $C=0.2$, for three different values of $p$. Recall that the onset of the steady mode is independent of both $p_m$ and $p$. Both $k_c^2$ (for steady and oscillatory modes) and $\omega_c^2$ are increasing functions of $Q$. The overall critical Rayleigh number $R_c$ is also an increasing function of $Q$, with this being more pronounced for the steady mode. At sufficiently large $Q$, oscillatory convection is preferred, with the value of $Q$ at the transition decreasing with increasing $p$. The transition value does however become independent of $p$ for $p$ sufficiently large. Assuming that $k_c^2$, $R$ and $s$ are $O(1)$ as $p \to \infty$, as may be verified \textit{ex post facto}, the fourth order dispersion relation~\eqref{eq:dr3} reduces to the quadratic equation
\begin{equation}
2C (\beta^4+Q) s^2 + \left( (\beta^4+Q) - 2 C R k^2 \right) s + \beta^2 (\beta^4+Q) - R k^2 = 0.
\label{eq:dr3simple}
\end{equation}
The other two roots of \eqref{eq:dr3} have large negative real parts. It should also be noted that in this large $p$ limit, expression~\eqref{eq:dr3simple} holds not just for $p_m=0$, but is in fact valid for any $O(1)$ value of $p_m$.
Bifurcation to oscillatory convection occurs when 
\begin{equation}
R^{(o)} = \frac{\beta^4 +Q}{2 C k^2} \qquad
\textrm{with} \qquad
 \omega^2 = \frac{\beta^2}{2C} - \frac{1}{4C^2}.
\end{equation}
Minimising $R^{(o)}$ with respect to $k^2$ gives
\begin{equation}
R_c^{(o)} = \frac{1+\left(1+Q\right)^{1/2}}{C} \quad \textrm{with} \quad
k_c^2 = \left(1+Q\right)^{1/2} \quad \textrm{and} \quad
\omega_c^2 = \frac{1+\left(1+Q\right)^{1/2}}{2C} - \frac{1}{4C^2}.
\label{eq:large_p_asymp}
\end{equation}
Although these results are derived in the limit of $p \to \infty$, they are remarkably accurate even for values as low as $p=5$, as can be seen from figure~\ref{fig:fig2}.

In many problems of this type, it is very challenging to find any analytical result determining the transition between two modes with unrelated wavenumbers. In the present case, however, we have a formula, albeit implicit, for $R_c^{(s)}$ \cite{WP_2014},
\begin{equation}
Q = R_c^{(s)} - 3 \left( \frac{R_c^{(s)}}{2}\right)^{2/3}.
\label{eq:WP}
\end{equation}
For $p \gg 1$ we also have an explicit relation for $R^{(o)}$, given by \eqref{eq:large_p_asymp}. At the transition between steady and oscillatory modes, $R_c^{(s)} = R_c^{(o)}$, leading to the following implicit expression for $Q$ in terms of $C$:
\begin{equation}
\label{eq:Q_transition}
Q = \frac{1+\left(1+Q\right)^{1/2}}{C} -
3 \left( \frac{1+\left(1+Q\right)^{1/2}}{2C} \right)^{2/3}.
\end{equation}
Note that at $Q=0$, $C=8/27$, consistent with \eqref{eq:C827}. Hence, for $C>8/27$, and sufficiently large $p$, oscillatory convection is preferred for all values of $Q$. We note also from \eqref{eq:Q_transition} that for $Q \gg 1$, the transition value of $C$ is very small, $C=O(Q^{-1/2})$; we shall discuss this regime in detail in the following section.

\begin{figure}[!h]
\includegraphics[trim=0cm 0 0 0cm, width=5.2in]{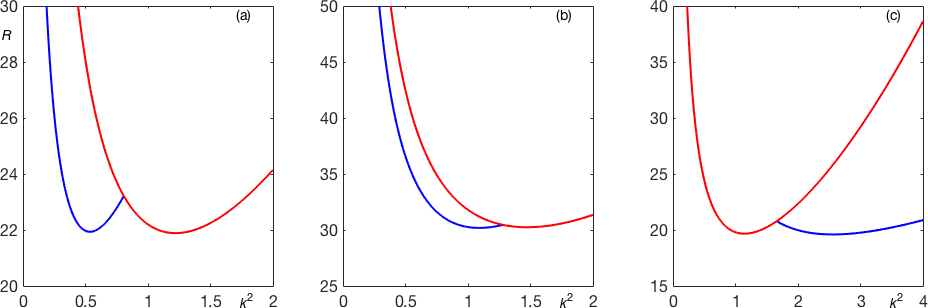}
\caption{Rayleigh numbers $R$ at the onset of instability as a function of $k^2$, for steady convection (red) and oscillatory convection (blue) at the value of $Q$ for which the minima of the steady and oscillatory curves are equal: $C=0.2$, $p_m=0.5$ in all plots, with (a)~$p=0.1$, $Q=7.1$; (b)~$p=0.5$, $Q=11.9$; (c)~$p=5$, $Q=5.9$.}
\label{fig:fig3}
\end{figure}

We now consider the case of finite $p_m$. We note that for $p \lesssim p_m$, the transition from steady to oscillatory onset as $Q$ is increased occurs with a decrease in wavenumber. This is illustrated in figure~\ref{fig:fig3}, which shows a typical transition for $p/p_m$ small, unity and large. For small $p$ we can make analytical progress in describing the transition, using methods similar to those employed above for large $p$. From equations~\eqref{eq:omega} and \eqref{eq:omega_two_exp}, we obtain, at leading order,
\begin{equation}
\omega^2 = \frac{p^2}{p_m^2} \left(p_mQ-\beta^4 \right), \qquad
R^{(o)} = \left(\frac{1+p_m}{p_m}\right) \frac{\beta^6}{k^2}.
\label{eq:small_p}
\end{equation}
Minimising $R^{(o)}$ over $k^2$ gives the values
\begin{equation}
k^2 = \frac{1}{2}, \qquad 
\omega^2 = \frac{p^2}{p_m^2} \left(p_mQ-\frac{9}{4} \right), \qquad
R^{(o)} = \frac{27}{4} \left(\frac{1+p_m}{p_m}\right).
\label{eq:small_p_asymp}
\end{equation}
Although the expression for $R^{(o)}$ does not contain $Q$, it should be noted that oscillations are possible only if $\omega^2 >0$ for some $\beta$, i.e.\ $Q > 1/p_m$. Combining expression~\eqref{eq:WP} with \eqref{eq:small_p_asymp}
shows that oscillations are preferred when
\begin{equation}
Q > \frac{27}{4} \left(\frac{1+p_m}{p_m}\right)^{2/3} \left[ \left(\frac{1+p_m}{p_m}\right)^{1/3}-1 \right].
\label{eq:small_p_transition}
\end{equation}
It is interesting to note that this is independent of $C$. From \eqref{eq:small_p_asymp} it can be seen that for $\omega^2$ to be positive at the transition we must have $Q> 9/(4p_m)$; this is always the case when \eqref{eq:small_p_transition} is satisfied.

\begin{figure}[!h]
\centering\includegraphics[trim=1cm 0 0 0cm, width=5in]{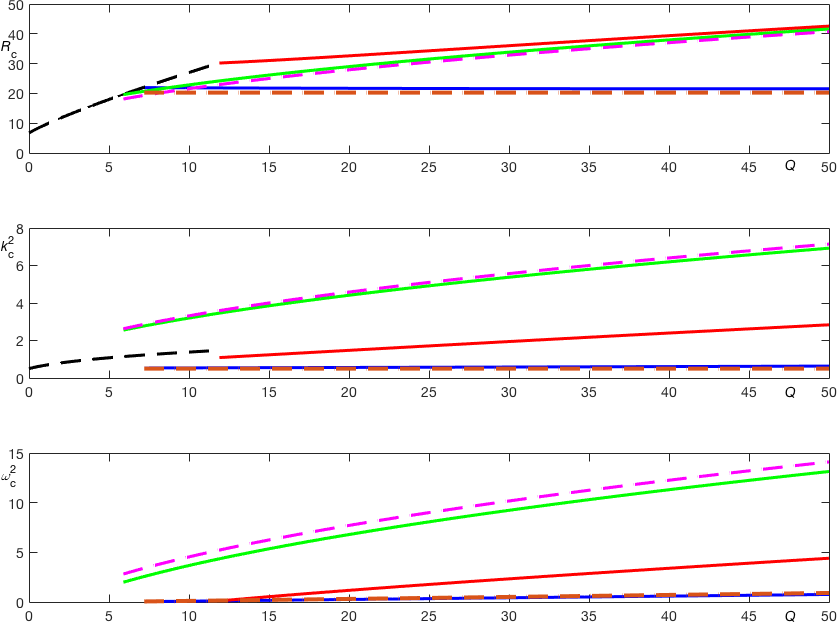}
\caption{$R_c$, $k_c^2$ and $\omega_c^2$ as functions of $Q$ for $p_m=0.5$ and $C=0.2$, with $p=0.1$ (blue), $p=0.5$ (red), $p=5$ (green). Bifurcation to oscillatory convection is shown as solid lines; the onset of steady convection, which is independent of $p$, is depicted by a dashed black line. The magenta dashed lines show the asymptotic results~\eqref{eq:large_p_asymp} for large $p$; the orange dashed lines show the asymptotic results~\eqref{eq:small_p_asymp} for small $p$.}
\label{fig:fig4}
\end{figure}

Figure~\ref{fig:fig4} shows examples of the change of critical mode from stationary to oscillatory as $Q$ is increased, for $p_m=0.5$, again with $C=0.2$ and with the three values of $p$ shown in figure~\ref{fig:fig3}. We have also shown the asymptotic results~\eqref{eq:small_p_asymp} for small $p$; these are extremely accurate even for $p=0.1$. It is interesting to note that the fact that the wavelength of the preferred oscillatory mode may be less than or greater than that of the steady mode, as illustrated by figure~\ref{fig:fig3}, leads to complicated dependence of the transition on $p$. In particular, for the parameter values adopted in figure~\ref{fig:fig4}, the 
transition value of $Q$ for $p = p_m$ is larger than that for both small and large values of $p$. As already noted, the large $p$ behaviour is independent of $p_m$, and so the $p=5$ results in figure~\ref{fig:fig4} are essentially identical to those in figure~\ref{fig:fig2}.

Finally, in this section, we consider the regime with $p_m \gg 1$. Oscillatory solutions are favoured at onset. On defining $\Lambda = Q/p_m$, expression~\eqref{eq:omega} becomes, at leading order, 
\begin{equation}
4C^2p\omega^4+\omega^2(1+p-2pC\beta^2)-p^2 \Lambda=0.
\label{eq:freq_large_pm}
\end{equation}
It is instructive to consider the case of $\Lambda \ll 1$; given that we are also considering the regime $p_m \gg 1$, this is not too restrictive an assumption --- the small $\Lambda$ regime still encompasses large values of $Q$. For small $\Lambda$, the two roots of equation~\eqref{eq:freq_large_pm} are given, at leading order, by
\begin{equation}
\omega^2 = \frac{p^2 \Lambda}{1+p-2pC\beta^2} \qquad \textrm{and} \qquad
\omega^2 = \frac{-(1+p-2pC\beta^2)}{4C^2p}.
\label{eq:freqs_large_pm}
\end{equation}
The smaller ($O(\Lambda)$) root for $\omega^2$ leads to
\begin{equation}
R^{(o)} = \frac{\beta^6}{k^2}, \quad \textrm{which is minimised when} \quad k^2 = \frac{1}{2}, \quad
\textrm{giving} \quad R^{(o)} = \frac{27}{4};
\end{equation}
this can occur only when $C<(1+p)/3p$. Note that when $Q = \Lambda =0$, the mode with $k^2=1/2$, $R=27/4$ is steady, with two of the roots of the dispersion relation being $s=0$ and $s=-p\beta^2/p_m$; as $Q$ is increased, the oscillatory roots arise from the merger and splitting of these two roots.

\begin{figure}[!h]
\centering\includegraphics[trim=1cm 0 0 0cm, width=5.1in]{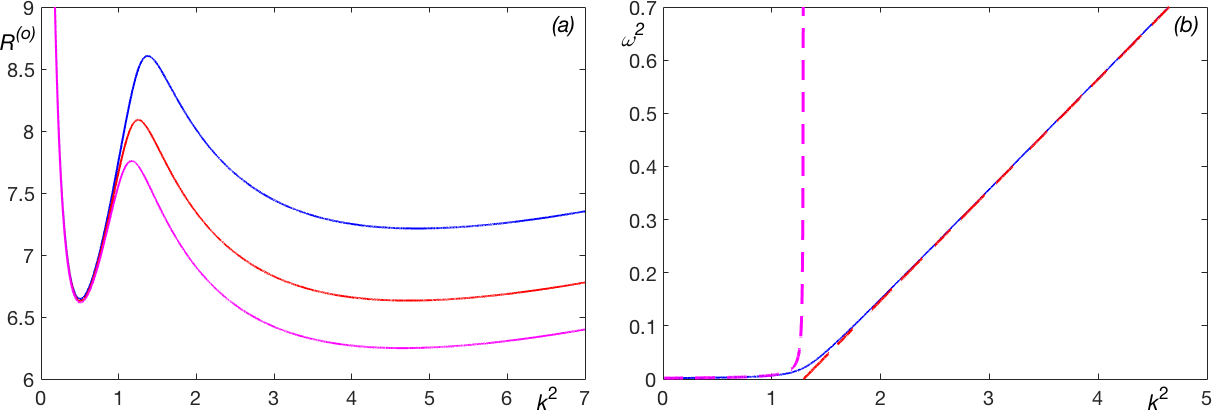}
\caption{$(a)$\ $R^{(o)}$ versus $k^2$ for $p_m=10^8$, $Q=10^7$, $p=0.1$, with $C=2.4$ (blue), $C=2.515$ (red), $C=2.6$ (magenta). $(b)$\ $\omega^2$ versus $k^2$ for the case of $C=2.4$ (solid line), together with the asymptotic results \eqref{eq:freqs_large_pm} (dashed lines).}
\label{fig:fig5}
\end{figure}

The larger ($O(1)$) root, which is that found when $Q=0$, is given at leading order by
\begin{equation}
R^{(o)} = \frac{1}{4C^2p^2} \left( (1+p) \frac{\beta^2}{k^2} + 2 Cp^2 \frac{\beta^4}{k^2} \right), 
 \quad \textrm{which is minimised when} \quad 
 k^2 = \sqrt{1+\frac{1+p}{2Cp^2}}.
\end{equation}
The condition that $\omega^2>0$ at this value of $k^2$ leads to the condition $C>(1+p)/(2+4p)$. Thus there are two minima of $R^{(o)}$ for values of $C$ in the range
\begin{equation}
\frac{1+p}{2(1+2p)} < C < \frac{1+p}{3p};
\label{eq:C_range}
\end{equation}
outside this range there is only one minimum. Figure~\ref{fig:fig5} shows $R^{(o)}$ versus $k^2$ for three values of $C$ in the range~\eqref{eq:C_range} for $Q=10^7$, $\Lambda=0.1$, showing cases where either the small or large $k^2$ mode is preferred, together with the transition value of $C$ where the two minima are equal. Also plotted is $\omega^2$ versus $k^2$ for one of the cases (all three have very similar frequencies), together with the small $\Lambda$ asymptotic results~\eqref{eq:freqs_large_pm}; it can be seen that except for a small range of $k^2$ (where $\beta^2 \approx (1+p)/2Cp$), the asymptotic results produce excellent agreement with the full system.

\section{$Q \gg 1$}

The case of very strong imposed fields ($Q \gg 1$) is of particular interest since it highlights how little Maxwell-Cattaneo effect is needed in order to bring about a qualitative change in the nature of the stability problem. It is, furthermore, a regime that allows analytical progress, thereby complementing the numerical approach, which, for general parameter values, is unavoidable.

As already noted, the onset of steady convection is given by expression~\eqref{eq:steady}, for all values of $C$. For the case of $Q \gg 1$, $R^{(s)}$ is minimised when, at leading order,
\begin{equation}
k^2 = \left( \frac{Q}{2} \right)^{1/3}, \quad \textrm{with} \quad R_c^{(s)} = Q.
\label{eq:Q_large_steady}
\end{equation}

When $C=0$, the problem reduces to that of classical magnetoconvection \cite{WP_2014}. In this case, if $p>p_m$ then the onset of instability can occur only via a steady bifurcation, and hence $R_c = R_c^{(s)}$. However, if $p<p_m$, then, for $Q \gg 1$, the onset of oscillatory instability occurs with the critical wavenumber and critical Rayleigh number given, at leading order, by
\begin{equation}
k^2 = \left( \frac{p^2}{(1+p)(p_m+p)}\right)^{1/3} \left( \frac{Q}{2}\right)^{1/3}, 
\qquad R_c^{(o)} = \frac{p^2(1+p_m)}{p_m^2(1+p)} Q.
\label{eq:Q_large_osc}
\end{equation}
It follows, from consideration of expressions~\eqref{eq:Q_large_steady} and \eqref{eq:Q_large_osc}, that if $p<p_m$ then $R_c^{(o)} < R_c^{(s)}$; i.e., oscillatory instability is always favoured in the regime of sufficiently large $Q$.

We might then ask, for $Q \gg 1$, how large does $C$ need to be to influence the results at leading order? As already noted, finite values of $C$ have no influence on the onset of steady convection; it is thus only the oscillatory modes that we need to consider. Inspection of the coefficients in the dispersion relation~\eqref{eq:dr3} --- for example, expression~\eqref{eq:a2} for $a_2$ --- suggests that $C$ will come into play when $CQ \sim \beta^2$, i.e.\ when $C \sim Q^{-2/3}$. However, this is misleading; instead, one should consider expression~\eqref{eq:omega} for the frequency at the onset of instability. When $k^2 \sim \beta^2 \sim Q^{1/3}$ and $C \sim Q^{-2/3}$, the leading order terms in \eqref{eq:omega} give rise to the expression:
\begin{equation}
4 C^2 p p_m^2 \omega^4 + p_m^2 (1+p) \omega^2 + p^2 (p-p_m)Q = 0.
\end{equation}
The $C=0$ root for $\omega^2$ is unchanged at this order, while the other root has $\omega^2<0$; thus with $C=O(Q^{-2/3})$ there is no leading-order influence of $C$ on the frequency of the marginal mode. The higher-order corrections to $\omega^2$ and $R_c$, which are both $O(Q^{2/3})$, can be calculated after some algebra, but their expressions are not particularly revealing.

The MC effect enters at leading order when $C = O(Q^{-1/2})$, leading to a set of scalings that differ from those of classical magnetoconvection. In this new regime, $k^2 \sim \beta^2 \sim Q^{1/2}$ and the coefficients in expression~\eqref{eq:omega} for the marginal frequency have magnitudes: $b_4 = O( Q^{-1})$, $b_2 = O(1)$, $b_0 = O(Q)$. Hence $\omega^2 = O(Q)$ for both roots of \eqref{eq:omega}, although only one root is meaningful with $\omega^2>0$. In this regime it becomes possible to scale the growth rate $s$ so as to remove all of the $Q$ dependence. On writing
\begin{equation}
s = Q^{1/2} s_0, \ C= Q^{-1/2} C_0, \ k^2= k_0^2 Q^{1/2}, \ \beta^2= \beta_0^2 Q^{1/2}, \ R = R_0 Q,
\label{eq:scaling}
\end{equation}
the coefficients of the quartic dispersion relation become, at leading order,
\begin{equation}
a_4 = 2C_0 p_m,
\end{equation}
\begin{equation}
a_3 = p_m + 2C_0 p (1+p_m) k_0^2 ,
\end{equation}
\begin{equation}
a_2 = (p+p_m+p p_m) k_0^2 + 2 C_0 p^2 (k_0^4 + 1) - 2C_0 p p_m R_0,
\end{equation}
\begin{equation}
a_1 =  (p + p^2 + p p_m) k_0^4 + p^2 - p p_m R_0 - 2C_0 p^2 R_0 k_0^2 ,
\end{equation}
\begin{equation}
a_0 =  p^2 (k_0^6 + k_0^2 - R_0 k_0^2).
\label{eq:a0_scaled}
\end{equation}
Note that we have almost recovered the full problem, described by \eqref{eq:dr3} -- \eqref{eq:a0}, with $Q=1$, but where we have now identified $k^2$ and $\beta^2$. It is thus straightforward, numerically, to explore the regime of $Q \gg 1$, $C=C_0Q^{-1/2}$, $C_0=O(1)$.

\begin{figure}[!h]
\includegraphics[trim=2cm 0 0 0cm, width=5.4in]{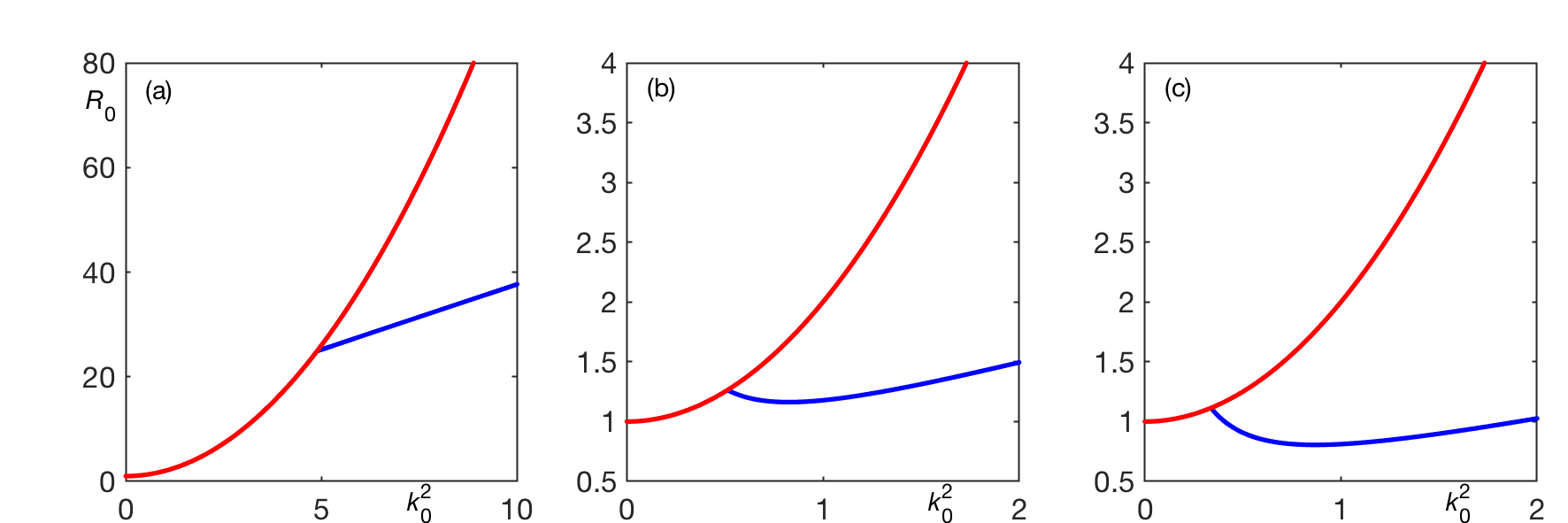}
\caption{Rayleigh numbers $R_0$ at the onset of instability as a function of $k_0^2$, for steady convection (red) and oscillatory convection (blue); $p=1$, $p_m=0.1$, with (a) $C_0=0.2$, (b) $C_0=1.1$, (c) $C_0=1.5$.}
\label{fig:fig6}
\end{figure}

\begin{figure}[!h]
\includegraphics[trim=2cm 0 0 0cm, width=5.4in]{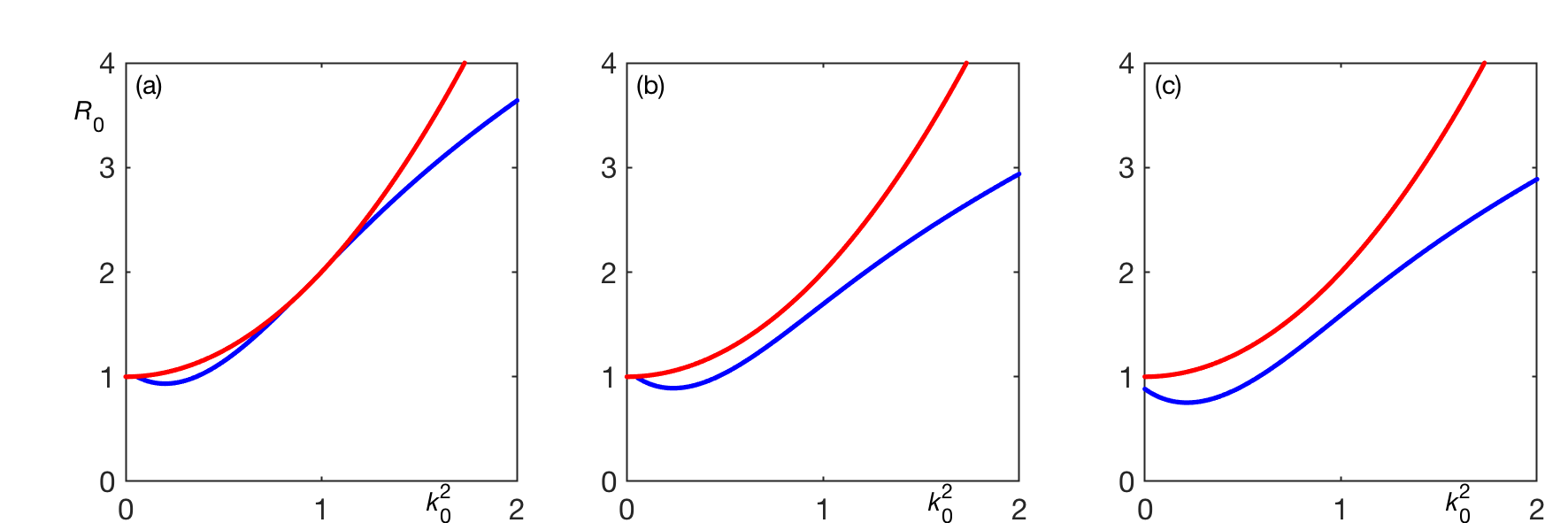}
\caption{Rayleigh numbers $R_0$ at the onset of instability as a function of $k_0^2$, for steady convection (red) and oscillatory convection (blue); $p=1$ for all plots; (a)~$p_m=0.95$, $C_0=0.51$; (b)~$p_m=0.95$, $C_0=0.6$; (c)~$p_m=1.1$, $C_0=0.6$.}
\label{fig:fig7}
\end{figure}

We should first note, from~\eqref{eq:a0_scaled}, that the critical value for a steady state bifurcation is given by
\begin{equation}
R_0^{(s)} = 1+k_0^4.
\end{equation}
The minimum value of $R_0$ is $R_0=1$, obtained when $k_0=0$; this is entirely consistent with the unscaled result~\eqref{eq:Q_large_steady}, which shows that the minimum value of $R$ is obtained when $k^2 = O(Q^{1/3})$, outside the present scaling.

Figures~\ref{fig:fig6} and \ref{fig:fig7} display the steady branch, together with possible different manifestations of the oscillatory branch as the parameters are varied. Figure~\ref{fig:fig6} shows $R_0^{(s)}$ and $R_0^{(o)}$ as functions of $k_0^2$ for the fixed values $p=1$, $p_m=0.1$, for three values of $C_0$. For sufficiently small $C_0$ (illustrated by $C_0=0.2$), the oscillatory branch appears at large wavenumber and increases with $k_0^2$; here the overall stability boundary is determined by the steady branch. As $C_0$ is increased (illustrated by $C_0=1.1$), the oscillatory branch develops a minimum, but this is still above the critical value for steady convection. A further increase in $C_0$ (illustrated by $C_0=1.5$) leads to the minimum of the oscillatory branch lying below $R_0=1$ and thus convection sets in as oscillations as $R_0$ is increased. Further structure can be identified by careful choices of $p_m$ and $C_0$, as shown in figure~\ref{fig:fig7}, in which we again fix $p=1$. In figure~\ref{fig:fig7}$a$, with $p_m=0.95$ and $C_0=0.51$, there is an isolated small region of wavenumbers where oscillations are possible, distinct from the branch at larger wavenumbers. For the example shown, the minimum of the oscillatory branch lies below the steady branch, and so defines the preferred mode, though it is possible to find parameter values for which this is not the case. On increasing $C_0$ slightly, to $C_0=0.6$ (figure~\ref{fig:fig7}$b$), the two oscillatory branches merge; topologically this is the same as that shown in figure~\ref{fig:fig6}$c$. On increasing $p_m$ to a value greater than $p$ (figure~\ref{fig:fig7}$c$ shows the case of $p_m=1.1$), the oscillatory branch lies wholly below the steady branch.

\begin{figure}[!h]
\includegraphics[trim=2.6cm 0 0 0cm, width=5.7in]{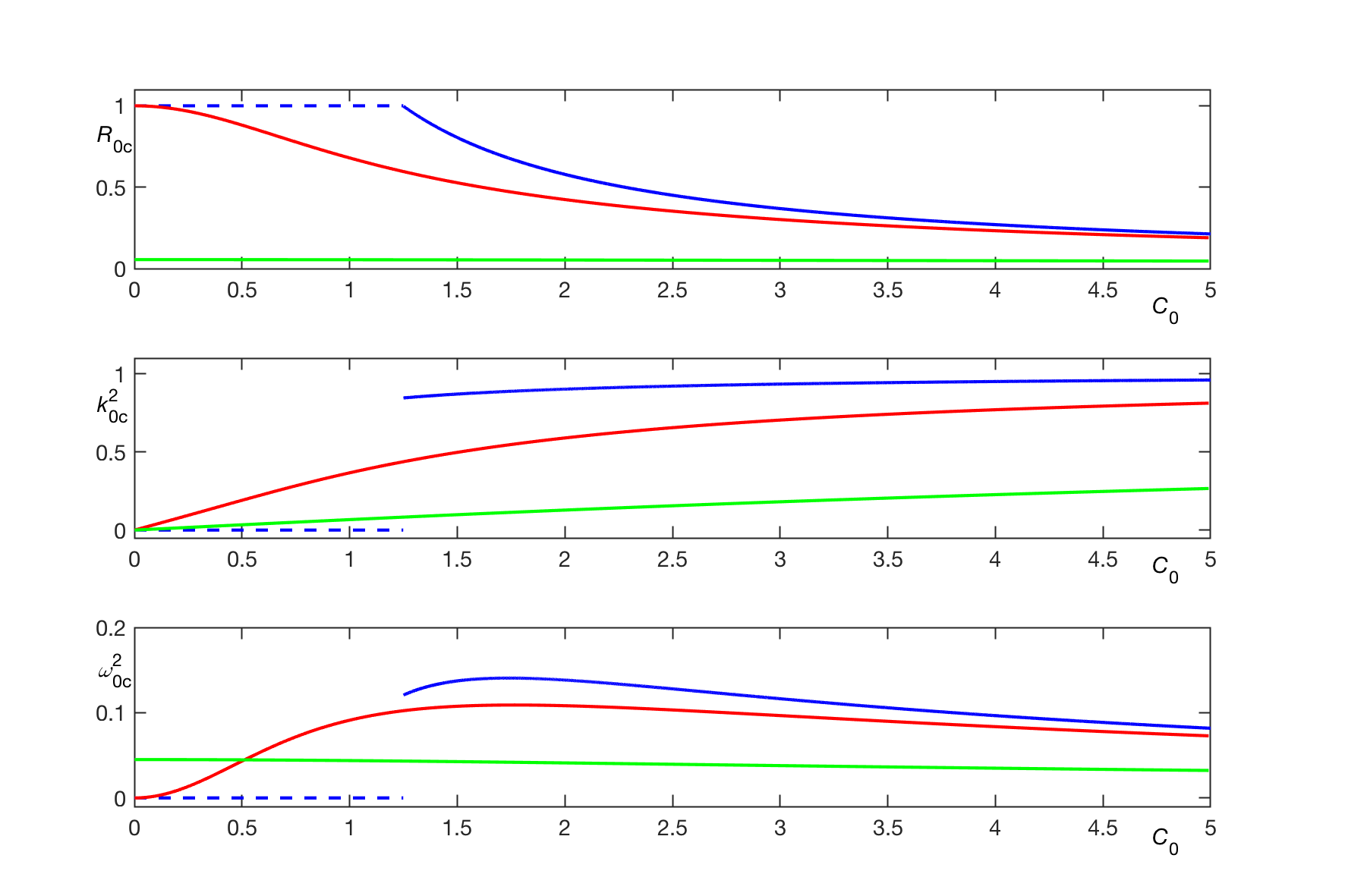}
\caption{$R_{0c}$, $k_{0c}^2$ and $\omega_{0c}^2$ 
governed by the scaling~\eqref{eq:scaling}. Here $p=1$, with $p_m=0.1$ (blue), $p_m=1$ (red), $p_m=10$ (green). Bifurcation to oscillatory (steady) convection is shown as a solid (dashed) line.}
\label{fig:fig8}
\end{figure}

Figure~\ref{fig:fig8} shows the critical values of $R_0$, $k_0^2$ and $\omega_0^2$ as functions of $C_0$ for $p=1$ and $\protect{p_m=0.1, \ 1, \ 10}$. When $p_m<p$, the preferred mode is steady for $C_0$ sufficiently small, but oscillatory for larger values. As $p_m$ increases, the range of $C_0$ for which the preferred mode is steady diminishes, vanishing when $p_m=p$. For $p_m>p$, the preferred mode is oscillatory for all $C_0$. It can be seen that on the oscillatory branches, for sufficiently large $C_0$, $R_{0c}$ and $\omega_{0c}^2$ decrease with $C_0$, whereas there is a gradual increase in $k_0^2$.

The trends shown in figure~\ref{fig:fig8} are consistent with the analysis of the critical curves for large $C_0$, i.e.\ large values of $C Q^{1/2}$. For such values of $C$, the preferred wavenumber continues to scale as $k^2 \sim Q^{1/2}$, with the coefficients of \eqref{eq:omega} given, at leading order, by 
\begin{equation}
 b_4 = 4 C^2 p p_m^2, \quad
b_2 = 4 C^2 p^3 ( k^4 + Q), \quad
b_0 = - 2 C p^3 (k^4+Q) k^2.
\end{equation}
One root has $\omega^2<0$. The other, meaningful, root is given, at leading order, by
\begin{equation}
\omega_c^2 = \frac{k^2}{2C} = \frac{Q^{1/2}}{2C},
\label{eq:omega_large_C}
\end{equation}
with
\begin{equation}
R_c = \frac{Q^{1/2}}{C}.
\label{eq:R_large_C}
\end{equation}
It is interesting to note that one can never escape the influence of the magnetic field if $Q \gg 1$, however large the value of $C$. The horizontal scale of the preferred mode is determined solely by $Q$, and hence is small; thus one can never recover the non-magnetic behaviour. This is a new mode of instability --- which may be classified as an MHD~MC mode --- for which both a strong magnetic field and the MC effect are vital. It is the dependence of $k_c^2$ on $Q$ that leads to the dependence of $\omega_c^2$ and $R_c$ on $Q$, shown in equations~\eqref{eq:omega_large_C} and \eqref{eq:R_large_C}.

The simplicity of expression~\eqref{eq:omega_large_C} leads us to ask whether this result could be obtained by making simplifying assumptions to the governing equations~\eqref{eq:goveq1}--\eqref{eq:goveq3}. Given the absence of the Prandtl number $p$ in \eqref{eq:omega_large_C} and \eqref{eq:R_large_C}, we are led to consider the limit of $p \to \infty$. On making the substitution $\tilde \calS = p \calS$ and letting $p \to \infty$ with $p_m$ finite, equation~\eqref{eq:goveq1} becomes
\begin{equation}
\label{eq:goveq1a}
0 = - R \theta + Q \frac{\partial}{\partial z} \left( \nabla^2 \tilde \calS \right) + \nabla^4 \calP,
\end{equation}
and equation~\eqref{eq:goveq3} becomes
\begin{equation}
\label{eq:goveq3a}
0 = \frac{\partial \calP}{\partial z} + \nabla^2 \tilde \calS.
\end{equation}
Equation~\eqref{eq:goveq2} is unchanged. We note that in \eqref{eq:goveq1a} the inertial term has been removed, and in \eqref{eq:goveq3a} the induction term no longer appears, equivalent to the limit of small magnetic Reynolds number. Elimination of $\tilde \calS$ between \eqref{eq:goveq1a} and \eqref{eq:goveq3a} leads to coupled equations for $\calP$ and $\theta$, with time derivatives appearing only in equation~\eqref{eq:goveq2}. This yields the dispersion relation~\eqref{eq:dr3simple}, with, as already noted,
\begin{equation}
R^{(o)} = \frac{\beta^4 +Q}{2 C k^2} \qquad
\textrm{with} \qquad
 \omega^2 = \frac{\beta^2 (\beta^4+Q) - R k^2}{2C (\beta^4+Q)} =
 \frac{\beta^2}{2C} - \frac{1}{4C^2}.
 \label{eq:omega_infp}
\end{equation}
We note that these expressions hold for all values of $Q$. For $Q \gg 1$, $R^{(o)}$ is minimised when $k^2 = Q^{1/2} $, with $R_c^{(o)} = Q^{1/2}/C$. The requirement that $\omega^2 > 0$ implies that $2C k^2 = 2C Q^{1/2} >1$. For large values of $C Q^{1/2}$, expression~\eqref{eq:omega_infp} reduces to \eqref{eq:omega_large_C}. We have thus identified the instability mechanism at large $C$ as one in which inertia and magnetic induction are unimportant.

Finally, in this section, we note, from inspection of the coefficients \eqref{eq:a2} -- \eqref{eq:a0}, that the regime of large $Q$ implicitly means the regime of large $p^2Q$; if $p$ is sufficiently small such that $p^2Q$ is not large, then the picture is quite different. As already noted in \eqref{eq:small_p}, in the limit of $p \to 0$, oscillatory modes are preferred for $Q$ sufficiently large, with 
\begin{equation}
R_c = R_c^{(o)} = \frac{27}{4} \frac{(1+p_m)}{p_m}.
\end{equation}
We note that there is no dependence on $C$, nor any explicit dependence on $Q$. For the problem of classical magnetoconvection ($C=0$), as $p$ is increased, $R_c$ first increases monotonically while $R_c = R_c^{(o)}$, before levelling off to its $p$-independent value once $R_c$ is determined by $R^{(s)}_c$. For non-zero $C$, however, the picture can be quite different. If we suppose that $p \ll 1$, but with $\Gamma = p^2 Q$ = $O(1)$, and assuming $k^2=O(1)$, $R^{(o)}$ may be approximated from equations \eqref{eq:sb_osc} -- \eqref{eq_c0} as
\begin{equation}
R^{(o)} = \frac{(1+p_m)}{p_m^3}
\left( 4 C^2 \Gamma^2 \frac{\beta^2}{k^2} - 4C p_m \Gamma \frac{\beta^4}{k^2} +
        p_m \Gamma \frac{\beta^2}{k^2} + p_m^2 \frac{\beta^6}{k^2} \right).
\label{eq:R_small_p}
\end{equation}
\begin{figure}[!h]
\includegraphics[trim=0cm 0 0 0cm, width=5.4in]{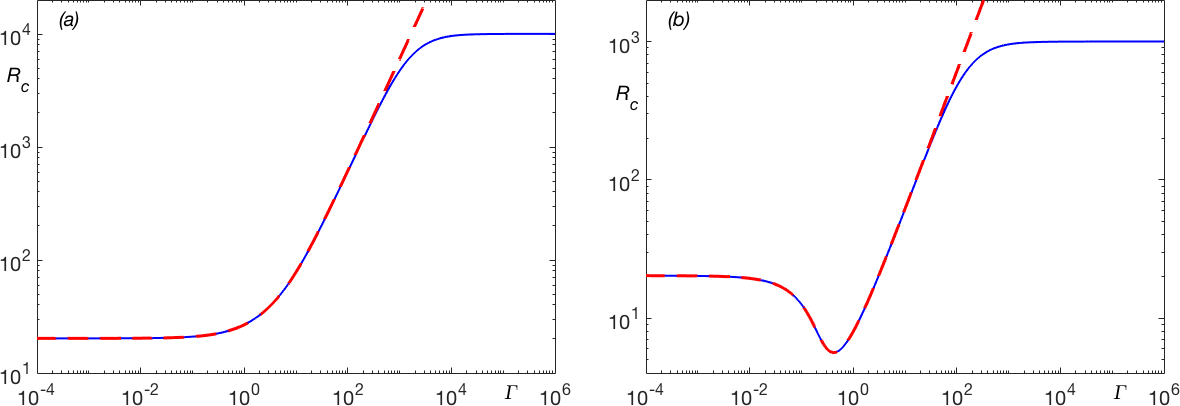}
\caption{$R_c$ as a function of $\Gamma = p^2Q$ for $p_m=0.5$, $Q=10^6$, and for $(a)$ $C=0.1$, $(b)$ $C=1$. The blue line is calculated from the full system; the red dashed line is calculated from the small $p$ result~\eqref{eq:R_small_p}.}
\label{fig:fig9}
\end{figure}
Figure~\ref{fig:fig9} plots $R_c$ (which here is given by $R^{(o)}$) as a function of $\Gamma$, computed from the full system and also from the asymptotic result~\eqref{eq:R_small_p}; the latter eventually fails once $\Gamma$ is no longer $O(1)$. As illustrated in Figure~\ref{fig:fig9}$a$, for small values of $C$, $R_c$ increases monotonically with $\Gamma$, as in classical magnetoconvection. However, as shown in Figure~\ref{fig:fig9}$b$, for larger values of $C$, there is a local minimum in $R_c$ at a finite value of $\Gamma$. Pleasingly, and somewhat surprisingly, it is possible to pin down the transition value of $C$ analytically.

We first note that on defining $\Gamma = p_m \gamma$, inspection of \eqref{eq:R_small_p} reveals that the transition value of $C$ is independent of $p_m$. To simplify the notation, we define $r=p_m R^{(o)}/(1+p_m)$; equation~\eqref{eq:R_small_p} then takes the form:
\begin{equation}
k^2 r=\beta^2(\Delta^2+\gamma), \quad \textrm{where} \quad \Delta=\beta^2-2C\gamma.
\label{eq:k2r}
\end{equation}
We wish to minimise $r$ over both $\gamma$ and $k^2$. First,
\begin{equation}
\frac{\partial r}{\partial \gamma} = 0 \implies \Delta = \frac{1}{4C},
\label{eq:minC_1}
\end{equation}
whereas
\begin{equation}
\frac{\partial r}{\partial k^2} = 0 \implies \Delta^2 + \gamma = 2 \beta^2 (\beta^2 - 1) \Delta.
\label{eq:minC_2}
\end{equation}
Eliminating $\Delta$ between \eqref{eq:minC_1} and \eqref{eq:minC_2} gives
\begin{equation}
1 + 16C^2 \gamma = 8C \beta^2 (\beta^2-1).
\label{eq:minC_3}
\end{equation}
Eliminating $\gamma$, using the expression in \eqref{eq:k2r}, together with \eqref{eq:minC_1}, leads to the expression
\begin{equation}
8C \beta^4 -16C \beta^2 +1 =0 \quad \textrm{or, equivalently,} \quad k^4 = 1 - \frac{1}{8C}.
\label{eq:minC_4}
\end{equation}
Thus there is a solution only if $C>1/8$. Finally, eliminating the wavenumber between \eqref{eq:minC_3} and \eqref{eq:minC_4} gives
\begin{equation}
\gamma=\frac{1}{2C}\left( 1+\sqrt{1-\frac{1}{8C}}-\frac{1}{4C}\right).
\end{equation}
For $\gamma>0$ we must therefore have
\begin{equation}
\sqrt{1-\frac{1}{8C}}>\frac{1}{4C}-1, \quad
\textrm{which reduces to} \quad C > \frac{1}{6}.
\end{equation}
Thus for $C>1/6$, $R_c$ is minimised at a finite value of $\Gamma$.

\section{Discussion}

The linear stability of a horizontal layer of viscous fluid permeated by a uniform vertical magnetic field has been studied in the presence of a uniform adverse temperature gradient. The heat flux-temperature relation is of non-Fourier type and in the linear regime takes the form discussed by \cite{Maxwell_1867,Cattaneo_1948}. The onset of instability is studied for the full range of Prandtl number $p$, magnetic Prandtl number $p_m$, Chandrasekhar number $Q$, and Maxwell-Cattaneo number $C$. While the onset of steady-state instability is unaffected by the Maxwell-Cattaneo (MC) effect, the new term in the heat equation makes a significant difference to the onset of oscillatory convection, particularly when $Q \gg 1$. This difference is manifested both in the critical Rayleigh number $R_c$ for the onset of convection and in the preferred wavenumber at which this critical value is attained. Furthermore, oscillatory convection may be found even when $p>p_m$, which is forbidden in classical magnetoconvection.

The calculations are most tractable (and informative) when $Q\gg 1$. It can be shown that the influence of the MC effect becomes significant when $C\sim Q^{-1/2}$, and indeed a reduced system of equations may be constructed in this distinguished limit that captures the effects of non-zero $C$ for its entire range, for the stress-free boundary case considered here. Crucially, the preferred wavenumber scales as $Q^{1/4}$, whereas if $C=0$ the preferred wavenumber is $O(Q^{1/6})$. This means that the Takens-Bogdanov bifurcation, where the oscillatory branch and the steady branch coincide, can be accommodated in the same scaling. Increasing $C$ in this regime at fixed $Q$ leads to a decrease in the critical value of $R$, which eventually scales like $Q^{1/2}/C$, while the critical wavenumber becomes independent of $C$. Interestingly, there can be up to three intersections of the steady and oscillatory branches. All possibilities can be found by varying the values of the parameters, as shown in figures~\ref{fig:fig6}, \ref{fig:fig7}. At large values of $p_m$, $R^{(o)}$ can have two minima in $k^2$ for the range of $C$ given by \eqref{eq:C_range}, with a transition of the preferred mode as $C$ is varied. For $p \ll 1$, $Q \gg 1$, with $\Gamma =  p^2 Q = O(1)$, the critical Rayleigh number has a non-monotonic dependence on $\Gamma$ --- in contrast to classical magnetoconvection --- provided that $C>1/6$.

It is important to discuss the astrophysical implications of the balance $C\sim Q^{-1/2}$, at which MC and magnetic effects both come into play. In terms of the lengthscale $d$ at which this balance is achieved, this relation may be expressed as
\begin{equation}
d\sim \frac{\tau_r p_m^{1/2} v_A}{p},
\label{eq:length_estimate}
\end{equation}
where $v_A = B/\sqrt{\mu_0 \rho}$ is the Alfv\'en speed. In the solar tachocline, for example, where $p_m =O(10^{-1})$, $p=O(10^{-6})$, $v_A =O(10^3 \mathrm{m}^2 \mathrm{s}^{-1})$, estimate~\eqref{eq:length_estimate} yields scales of the order of one metre. It is certainly conceivable that in astrophysical bodies in which convection occurs in the presence of very strong magnetic fields, as postulated, for example, in the dynamo phase of proto-neutron stars \cite{RG_2005}, the MC effect will be of importance over a much wider range of lengthscales. 

The results in this paper have been obtained with the simple stress-free mechanical boundary conditions and fixed boundary temperatures. What is the effect of varying the boundary conditions? The magnetic field conditions, while complicated in general, become unimportant in the large $Q$ case, since the reduced equations contain the magnetic field only through the quantity $\nabla^2{\cal S}$, which may thus be eliminated without differentiation. The thermal and mechanical boundary conditions will yield order unity differences in the critical values of the parameters. A detailed analysis shows, however, that when $Q$ is large, the dominant terms in the interior give the same solutions as in the stress-free, fixed temperature case, and the new boundary conditions have to be accommodated by boundary layers; in the case of small $C$ ($C \ll Q^{-2/3}$) there are Hartmann layers, of thickness O$(Q^{-1/2})$, and viscous and thermal layers with thicknesses O($Q^{-1/4}$) and O$(Q^{-1/6})$ respectively, the latter reflecting the size of the critical wavelength. These layers act to adjust the mainstream functions to the boundary conditions, but are passive at leading order, offering no change to the stress-free critical values. For larger $C$ ($C \geq Q^{-1/2}$), the critical wavenumber increases to O$(Q^{1/4})$ and so there are only two boundary layers, but they are still passive. Clearly, though, the higher order corrections to the critical values (not calculated here) will depend on the boundary conditions.

The calculations described in this paper could be developed in a variety of ways. Including nonlinear effects would enable us to see if the preferred form of oscillations would be stationary or travelling waves (in one horizontal dimension), or which of the many possible nonlinear modes could occur in the fully three-dimensional case. Such questions have been extensively investigated in the $C=0$ case by \cite{WP_2014}. The influence of MC effects on other double diffusive type problems (such as rotating convection or thermosolutal convection) have the potential to uncover new phenomena; we are currently preparing a paper on the latter problem.

\maketitle
%
%
%

\ack{DWH and MREP thank Sultan Qaboos University for hosting two visits during which part of this work was carried out; IAE also acknowledges that some of this work was carried out while he was at Sultan Qaboos University. \\
Funding. We are grateful to Sultan Qaboos University for providing financial support for research visits. DWH was also partially supported by STFC grant ST/N000765/1, and MREP by King's College, Cambridge.}


 \bibliographystyle{RS}
  \bibliography{refs_mcmc}
\end{document}